\documentclass[longauth,letter,traditabstract]{aa} 
\usepackage{graphicx}
\usepackage{natbib}
\usepackage{txfonts}
\begin{document}
   \title{A study of the distant activity of comet C/2006~W3~(Christensen)
using {\it Herschel} and ground-based radio telescopes
\thanks{{\it Herschel} is an ESA space observatory with science instruments
  provided by European-led Principal Investigator consortia and with important participation from NASA.}}

\author{D.~Bockel\'ee-Morvan\inst{1}
\and P.~Hartogh\inst{2} \and J.~Crovisier\inst{1} \and
B.~Vandenbussche\inst{3} \and B.M.~Swinyard\inst{4} \and
N.~Biver\inst{1} \and D.C.~Lis\inst{5} \and C.~Jarchow\inst{2}
\and R.~Moreno\inst{1} \and D.~Hutsem\'ekers\inst{6}  \and
E.~Jehin\inst{6} \and M.~K\"{u}ppers\inst{11} 
L.M.~Lara\inst{12}  \and E.~Lellouch\inst{1} \and J.~Manfroid\inst{6} \and M.~de~Val-Borro\inst{2}  \and
S.~Szutowicz\inst{7} \and M.~Banaszkiewicz\inst{7} \and
F.~Bensch\inst{8} \and M.I.~Blecka\inst{7} \and
M.~Emprechtinger\inst{5} \and T.~Encrenaz\inst{1} \and
T.~Fulton\inst{9}  \and
M.~Kidger\inst{10}   \and M.~Rengel\inst{2}
\and C.~Waelkens\inst{3} \and E.~Bergin\inst{13} \and
G.A.~Blake\inst{5} \and J.A.D.L.~Blommaert\inst{3} \and
J.~Cernicharo\inst{14} \and L.~Decin\inst{3} \and
P.~Encrenaz\inst{15} \and T.~de Graauw\inst{16,17,18} \and S.~Leeks\inst{4} \and A.S.~Medvedev\inst{2} \and
D.~Naylor\inst{19} \and R.~Schieder\inst{20} 
\and N.~Thomas\inst{21} }


\institute{LESIA, Observatoire de Paris, 5 place Jules Janssen, F-92195 Meudon, France\\ 
\email{dominique.bockelee@obspm.fr}
\and Max-Planck-Institut f\"ur Sonnensystemforschung, Katlenburg-Lindau, Germany           
\and Instituut voor Sterrenkunde, Katholieke Universiteit Leuven, Belgium                
\and Rutherford Appleton Laboratory, Oxfordshire, United Kingdom                         
\and California Institute of Technology, Pasadena, United States                         
\and F.R.S.-FNRS, Institut d'Astrophysique et de G\'eophysique, Li\`ege, Belgium                      
\and Space Research Centre, Polish Academy of Science, Warszawa, Poland                  
\and Deutsches Zentrum f\"{u}r Luft- und Raumfahrt (DLR), Bonn, Germany                  
\and Blue Sky Spectroscopy Inc., Lethbridge, Alberta, Canada                             
\and {\it Herschel} Science Centre, ESAC, Madrid, Spain                                       
\and European Space Astronomy Centre, Madrid, Spain                                     
\and Instituto de Astrof\'isica de Andaluc\'ia (CSIC), Granada, Spain                   
\and University of Michigan, Ann Arbor, United States                                   
\and CAB. INTA-CSIC Crta Torrejon a Ajalvir km 4. 28850 Torrejon
de Ardoz, Madrid, Spain                                                                 
\and LERMA, Observatoire de Paris, and Univ. Pierre et Marie Curie, Paris, France       
\and SRON Netherlands Institute for Space Research, PO Box 800, Groningen, 
the Netherlands                                                     
\and Leiden Observatory, University of Leiden, the Netherlands
\and Joint ALMA Observatory, Santiago, Chile
\and University of Lethbridge, Canada                                                   
\and University of Cologne, Germany                                                     
\and University of Bern, Switzerland                                                    
 }

   \date{7 May 2010; Accepted for A\&A Herschel Special Issue}

\abstract{Comet C/2006 W3 (Christensen) was observed in November
2009 at 3.3 AU from the Sun with {\it Herschel}.The PACS instrument
acquired images of the dust coma in 70-$\mu$m and 160-$\mu$m
filters, and spectra covering several H$_2$O rotational lines. Spectra
in the range 450--1550 GHz were acquired with SPIRE. The comet emission 
continuum from 70 to
672 $\mu$m was measured, but no lines were detected. The spectral
energy distribution indicates thermal emission from large
particles and provides a measure of the size distribution index
and dust production rate. The upper limit to the water production
rate is compared to the production rates of other species (CO,
CH$_3$OH, HCN, H$_2$S, OH) measured with the IRAM 30-m and
Nan\c{c}ay telescopes. The coma is found to be strongly enriched
in species more volatile than water, in comparison to comets
observed closer to the Sun. The CO to H$_2$O production rate ratio
exceeds 220\%. The dust to gas production rate ratio is on the
order of 1.

}


\authorrunning{Bockel\'ee-Morvan et al.}

\titlerunning{Comet C/2006 W3 (Christensen)}

\maketitle
%

\section{Introduction}
Direct imaging shows that distant activity is a general property
of cometary nuclei \citep[e.g.,][]{mazo09}. It is attributed to
the sublimation of hypervolatile ices, such as CO or
CO$_2$, or to the release of volatile species trapped in amorphous
water ice during the amorphous-to-crystalline phase transition
\citep{pria04}. Indeed, at heliocentric distances $r_h$ larger
than 3--4 AU, the sublimation of water, the major volatile in
cometary nuclei, is inefficient. Characterizing the processes
responsible for distant activity is important for understanding
the structure and composition of cometary nuclei, their thermal
properties and their evolution upon solar heating. However,
detailed investigations of distant nuclei are sparse. The
best studied objects are the distant comet
29P/Schwassmann-Wachmann 1, where CO, CO$^+$, and CN were detected
at 6 AU from the Sun, and C/1995 O1 (Hale-Bopp),
 whose exceptional activity allowed us to detect several 
molecules and radicals farther than 3 AU --- including CO up to 14
AU \citep{biv02,rauer03}.

Comet C/2006 W3 (Christensen) was discovered in November 2006 at
$r_h$ = 8.6 AU from the Sun with a total visual magnitude $m_v$
$\sim$18. This long-period comet passed
perihelion on 9 July 2009 at $r_h$ = 3.13 AU. Because of its
significant brightness ($m_v$ $\sim$ 8.5 at perihelion), it was an
interesting target for the study of distant cometary activity. We
report here on observations undertaken at $r_h$ =
3.3 AU post-perihelion with the PACS \citep{Poglitsch10} and SPIRE 
\citep{Griffin10} instruments on {\it Herschel} \citep{Pilbratt10}, in the
framework of the {\it Herschel} Guaranteed Time Key Project called
''Water and related chemistry in the Solar System'' \citep{hart09}.
These observations are complemented by production rate
measurements of several species using the Nan\c{c}ay radio
telescope and the 30-m telescope of Institut de Radioastronomie
millim\'etrique (IRAM) at 3.2--3.3 AU pre- and post-perihelion.


\section{Observations with {\it Herschel}}

Comet C/2006 W3 (Christensen) was observed with {\it Herschel} on 1--8
November, 2009 at $r_h$ $\sim$ 3.3 AU and a distance from {\it Herschel} 
$\Delta$ = 3.5--3.7 AU. 

\begin{figure}
\centering
\includegraphics[angle=270,width=9cm]{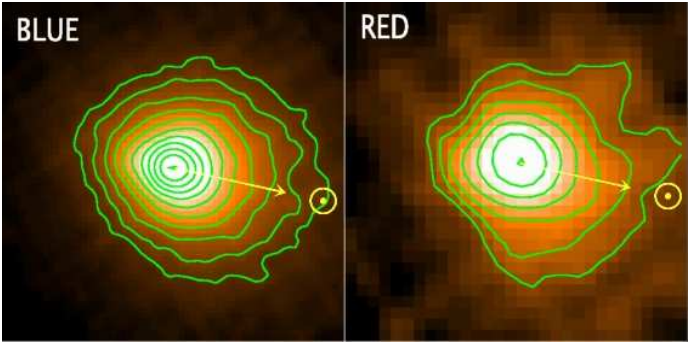}
\caption{Blue (70 $\mu$m) and Red (160 $\mu$m)
1\arcmin$\times$1\arcmin~maps of C/2006 W3 (Christensen) observed
with PACS on 1 November 2009 UT. Pixel size is 1\arcsec~in the
Blue map and 2\arcsec~in the Red map. Contours levels are stepped
by 0.1 in Log, up to 99\% of maximum intensity. East is on the
left, North is up. The Sun direction is indicated.}
              \label{Fig:pacsmap}
               \end{figure}

The PACS observations, acquired during the {\it Herschel} Science
Demonstration phase, consisted of: 1) on November 1.83 UT,
simultaneous acquisition of 8\arcmin$\times$11\arcmin~coma images
(Fig.~\ref{Fig:pacsmap}) in Blue (60-85 $\mu$m) and Red (130--210
$\mu$m) bands using the scan map photometry mode with a scan speed of
10\arcsec/sec (Obsid $\#$1342186621 and 1342186622 with orthogonal
scanning, duration $t_{int}$ = 565 s each), and 2) on November
8.74--8.82 UT, pointed source dedicated line spectroscopy over a
47\arcsec$\times$47\arcsec~field of view (5$\times$5  pixels of
9.4\arcsec) with a large (6\arcmin) chopper throw and a number
(so-called line repetition $l_{rep}$) of ABBA nodding cycles
(Obsid $\#$1342186633, $t_{int}$ = 6837 s). The water lines
2$_{21}$--1$_{10}$ (108.07 $\mu$m, $l_{rep}$=2),
3$_{13}$--2$_{02}$ (138.53 $\mu$m, $l_{rep}$=2),
3$_{03}$--2$_{12}$ (174.63 $\mu$m, $l_{rep}$=3),
2$_{12}$--1$_{01}$ (179.53 $\mu$m, $l_{rep}$=1) and
2$_{21}$--2$_{12}$ (180.49 $\mu$m, $l_{rep}$=1) were targeted at a
spectral resolution $\Delta \lambda$ $\sim$ 0.11--0.12 $\mu$m.
PACS data were processed with the HIPE software version 2.3.1
which uses ground calibration for the signal to flux conversion. A
high pass filter of width 297 was used to remove the $1/f$ noise.
According to sky calibration sources, PACS spectroscopy fluxes in
the 100--220 $\mu$m range are too high by a factor of 1.1 on average,
with a 30\% accuracy. For the Blue and Red maps, the
overestimation factors are 1.05 and 1.29, and the flux accuracies are 
10 and 20\%, respectively. We applied these corrections. No hint
of lines is present in the PACS spectra
($online$ Fig.~\ref{Fig:pacsspectra}, Table~\ref{table-I} for upper limits).
However, thermal emission from the dust coma is detected in the 25
 pixels.

\onlfig{2}{
\begin{figure*}
\includegraphics[width=13cm, angle=270]{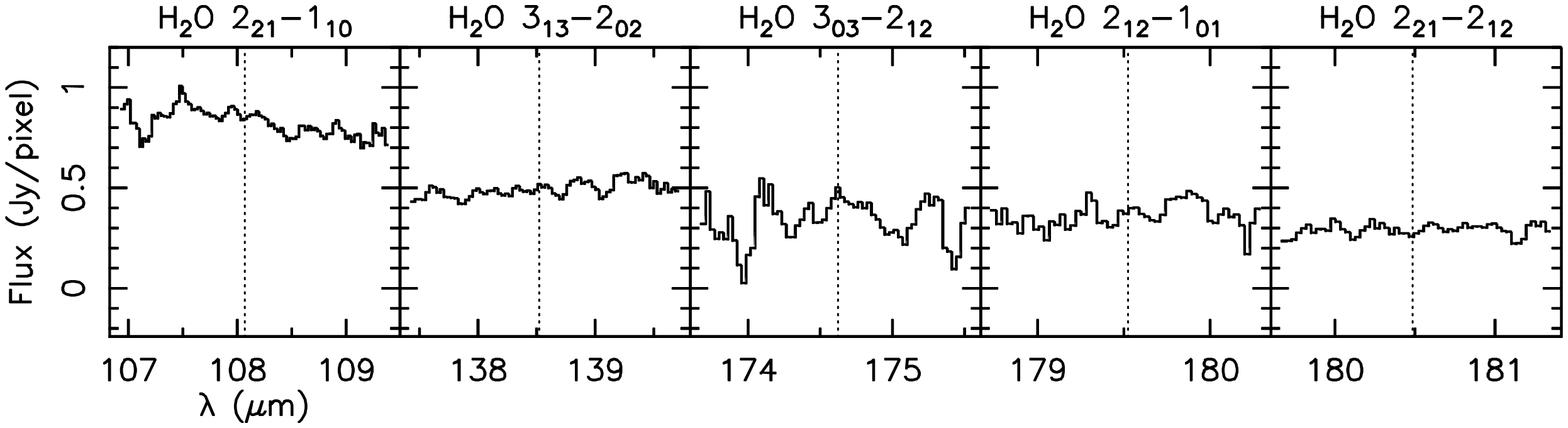}
\caption{Spectra of C/2006 W3 (Christensen) obtained with PACS on
8 November, 2009 (central pixel pointed on the comet nucleus).
The pixel size is 9.4\arcsec$\times$9.4\arcsec. The wavelengths of
water lines are indicated by vertical dashed lines.}
\label{Fig:pacsspectra}
\end{figure*}
}

\begin{figure}
\includegraphics[width=6.5cm,angle=270]{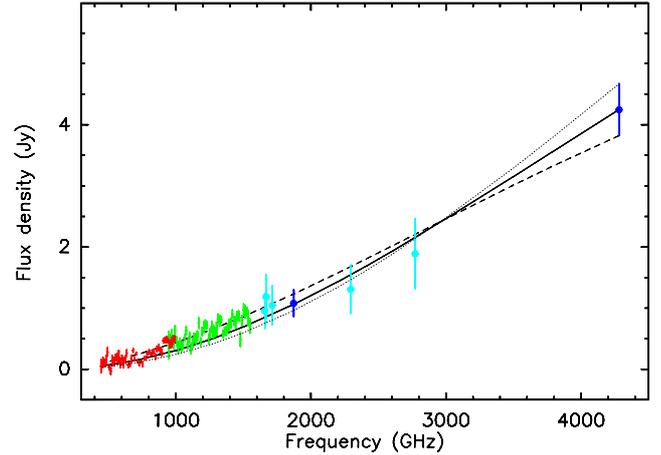}
\caption{Spectral Energy Distribution of C/2006 W3 (Christensen)
combining SPIRE (SLW: red dots; SSW: green dots), PACS photometry (blue dots), 
and spectroscopy (cyan dots) data,
scaled to a Half Power Beam Width of 18.7\arcsec~and $\Delta$=3.65 AU. 
SPIRE data were binned from $\sim$ 0.02 to 0.4 cm$^{-1}$. Curves are 
models for amorphous carbon grains with $a_{max}$= 0.9 mm, and 
size index --3.6 (solid), --2.8 (dashed), and --3.85 (dotted) (see text). }
              \label{Fig:spire-spectrum}%
    \end{figure}

On Nov. 6.59--6.68 UT, the comet was observed using the central
pixels on the SPIRE spectrometer for a total duration of 6926 s.
The spectral range (450--1550 GHz) encompasses the fundamental
ortho 1$_{10}$--1$_{01}$ and para 1$_{11}$--0$_{00}$ water lines.
This source is extremely faint for the SPIRE spectrometer. Standard 
pipeline
reduction shows significant problems with the overall flux level 
in both the high- (SSW) and low- (SLW) frequency channels. A
different approach has been taken therefore that does not rely on
the standard pipeline processing. It uses the variation in
bolometer temperature to transform the source and dark
interferograms into spectra which are then subtracted and divided
by a calibration spectrum obtained on a bright source (Uranus). 
The other
aspect that is not currently accurately calibrated in standard
processing is the effect of variations in the instrument
temperature between the observation of the dark sky and the
source: this can cause large relative variations in the recorded
spectrum in the low frequency SLW portion. For the data presented
here we have determined the overall net
flux of the source with no subtraction or addition of flux from
the variation in instrument temperature. We then inspect the low
frequency spectrum and compare to the spectrum expected from the
subtraction of two black bodies at temperatures given by the
average instrument temperatures recorded in the housekeeping data.
In general, as with the standard pipeline, this gives either too
much or too little flux and a non-physical spectrum compared to
the spectrum seen in the (unaffected) high frequency band. The
difference in model instrument temperatures in the dark sky and
the source observation are therefore varied until a match between
the overall flux level in SSW is achieved and the flux at the
lowest frequencies is zero within the noise.  The degree of
variation in the temperature difference required to achieve this
is less than 1\% of the recorded temperatures. The comet is
somewhat extended in the beam of the SPIRE spectrometer so a
correction is required to the shape of the spectrum to account for
the varying beam size of the instrument with frequency
\citep{swin10}. We have normalised this correction here to give
the flux in an effective beam size of 18.7\arcsec~for both the SSW
and SLW channels. Inspection of the data shows that no line is
detected within a 3-$\sigma$ upper limit of
0.7--0.8$\times$10$^{-17}$ W m$^{-2}$. Continuum emission from the
coma is detected (Fig.~\ref{Fig:spire-spectrum}).

\section{Observations with ground-based telescopes}

\begin{table*}
\caption[]{Molecular observations in comet C/2006~W3
(Christensen)} \label{table-I}
\begin{tabular}{lcclclcc}
\hline
\cline{1-8}\\[-0.2cm]
\phantom{IRAM00}UT date & $<r_{h}>$ & $<\Delta>$ & Line & Intensity    & Unit &Velocity shift &  Prod. rate\\[0.cm]
\phantom{IRAM00}[mm/dd.dd] & [AU]  & [AU]       &      &        &  & [km~s$^{-1}$]   &  [s$^{-1}$] \\
\cline{1-8}\\[-0.2cm]
PACS\phantom{00}11/8.74--8.82         &   3.35    &     3.70       & H$_2$O(2$_{12}$--1$_{01}$)$^{\footnotesize a}$ &  $<$ 2.2$\times$10$^{-18}$$^{\footnotesize a}$ & W m$^{-2}$  &  --  & $<$ $1.4\times10^{28}$\\[0cm]
\cline{1-8}\\[-0.2cm]
SPIRE\phantom{0}11/6.59--6.68  & 3.34    &  3.65    & H$_2$O(1$_{11}$--0$_{00}$) &  $<$ 8$\times$10$^{-18}$  & W m$^{-2}$ &   --            & $<$ $4\times10^{28}$ \\[0cm]
\cline{1-8}\\[-0.2cm]
Nan\c{c}ay\phantom{0}02/10--04/19 & 3.27 & 3.90  & OH 18 cm   & $-0.016\pm0.003$  &   Jy~km~s$^{-1}$   & $-0.32\pm0.12$  & $3.8\pm0.9\times10^{28}$ \\
\cline{1-8}\\[-0.2cm]
IRAM\phantom{00}09/13.88 & 3.20 & 2.57 & HCN(1--0)  & \phantom{$-$}$0.271\pm0.015$ & K~km~s$^{-1}$$^{\footnotesize b}$  & $-0.21\pm0.03$ & $1.6\pm0.1\times10^{26}$\\
\phantom{IRAM00}09/13.93 & 3.20 & 2.57  & HCN(3--2)  & \phantom{$-$}$0.961\pm0.140$ & &$-0.15\pm0.06$ & $1.8\pm0.3\times10^{26}$\\
\phantom{IRAM00}09/14.86 & 3.20 & 2.59  & HCN(1--0)  & \phantom{$-$}$0.262\pm0.013$ & &$-0.10\pm0.03$ & $1.6\pm0.1\times10^{26}$\\
\phantom{IRAM00}09/14.93 & 3.20 & 2.59  & HCN(3--2)  & \phantom{$-$}$0.972\pm0.097$ & &$+0.03\pm0.04$ & $1.6\pm0.2\times10^{26}$\\
\phantom{IRAM00}09/14.88 & 3.20 & 2.59  & CO(2--1)   & \phantom{$-$}$0.433\pm0.028$ & &$-0.04\pm0.03$ & $3.9\pm0.3\times10^{28}$ \\
\phantom{IRAM00}09/14.40  & 3.20 & 2.58 &  CH$_3$OH($1_0$--$1_{-1}$E)$^{\footnotesize c}$ & \phantom{$-$}$0.086\pm0.011$ & &$-0.10\pm0.04$ & $1.5\pm0.3\times10^{27}$ \\
\phantom{IRAM00}09/14.93 & 3.20 & 2.59  & H$_2$S($1_{10}$--1$_{01}$)    & \phantom{$-$}$0.311\pm0.031$ & &$-0.05\pm0.04$ & $1.0\pm0.1\times10^{27}$ \\
\phantom{IRAM00}09/14.95 & 3.20 & 2.59  & CS(2--1)   & \phantom{$-$}$0.028\pm0.012$ &               & -- &$0.5\pm0.2\times10^{26}$ \\
\phantom{IRAM00}09/14.95 & 3.20 & 2.59 &  H$_2$CO($3_{12}$--$2_{11}$)   & $<0.203$ &   & -- &$<1\times10^{27}$ \\
\phantom{IRAM00}10/29.71 & 3.32 & 3.48  & HCN(1--0)  & \phantom{$-$}$0.165\pm0.010$ & &\phantom{$-$}$-0.15\pm0.03$ & $1.4\pm0.1\times10^{26}$\\
\phantom{IRAM00}10/29.71 & 3.32 & 3.48  & CO(2--1)   & \phantom{$-$}$0.248\pm0.026$ & &\phantom{$-$}$-0.11\pm0.03$ & $3.0\pm0.3\times10^{28}$ \\
\hline
\end{tabular}

$^{\footnotesize a}$ 3-$\sigma$ upper limit per pixel in the
central 9.4\arcsec$\times$ 9.4\arcsec pixel. The upper limit on the
flux per beam is estimated to be $\sim$2 times higher. Upper limits for H$_2$O
2$_{21}$--1$_{10}$, 3$_{13}$--2$_{02}$, 2$_{21}$--2$_{12}$, and
3$_{03}$--2$_{12}$ lines are 4.9, 1.8, 1.0, and 3.9
($\times$10$^{-18}$) W m$^{-2}$ respectively.

$^{\footnotesize b}$ For IRAM data, the intensity scale is the main beam brightness temperature. 

 $^{\footnotesize c}$ CH$_3$OH
$2_0-2_{-1}$E, $3_0-3_{-1}$E, $4_0-4_{-1}$E, $5_0-5_{-1}$E, and
$6_0-6_{-1}$E lines were also observed with line intensities of
$0.083$, $0.062$, $0.049$, $0.043$, and $0.022$ ($\pm0.010$) K km
s$^{-1}$, respectively.

\end{table*}

\begin{figure}[ht]
\centering
\includegraphics[height=\hsize,angle=270]{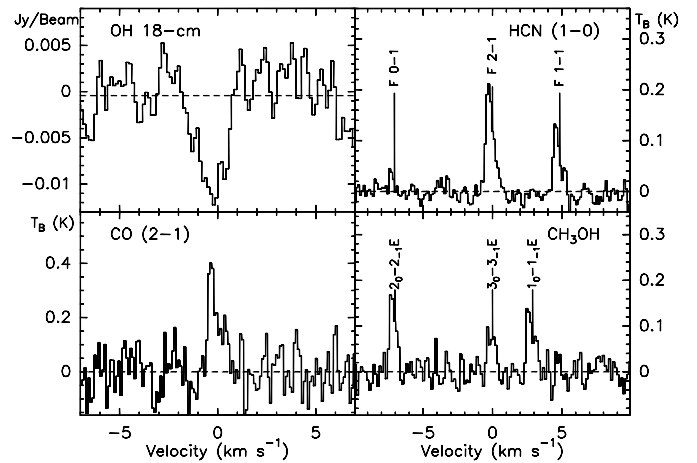}
\caption{Sample of ground-based spectra of comet C/2006 W3 (Christensen): OH
spectrum (average of 1665 and 1667~MHz lines) observed at
Nan\c{c}ay on Feb. 2 to Apr. 19, and HCN (Sept. 13-14), CO (Oct.
29), and CH$_3$OH (Oct. 14) spectra observed at the IRAM-30m.}
\label{Fig:nancay-iram}
\end{figure}

\onlfig{5}{
\begin{figure*}
\includegraphics[width=13cm]{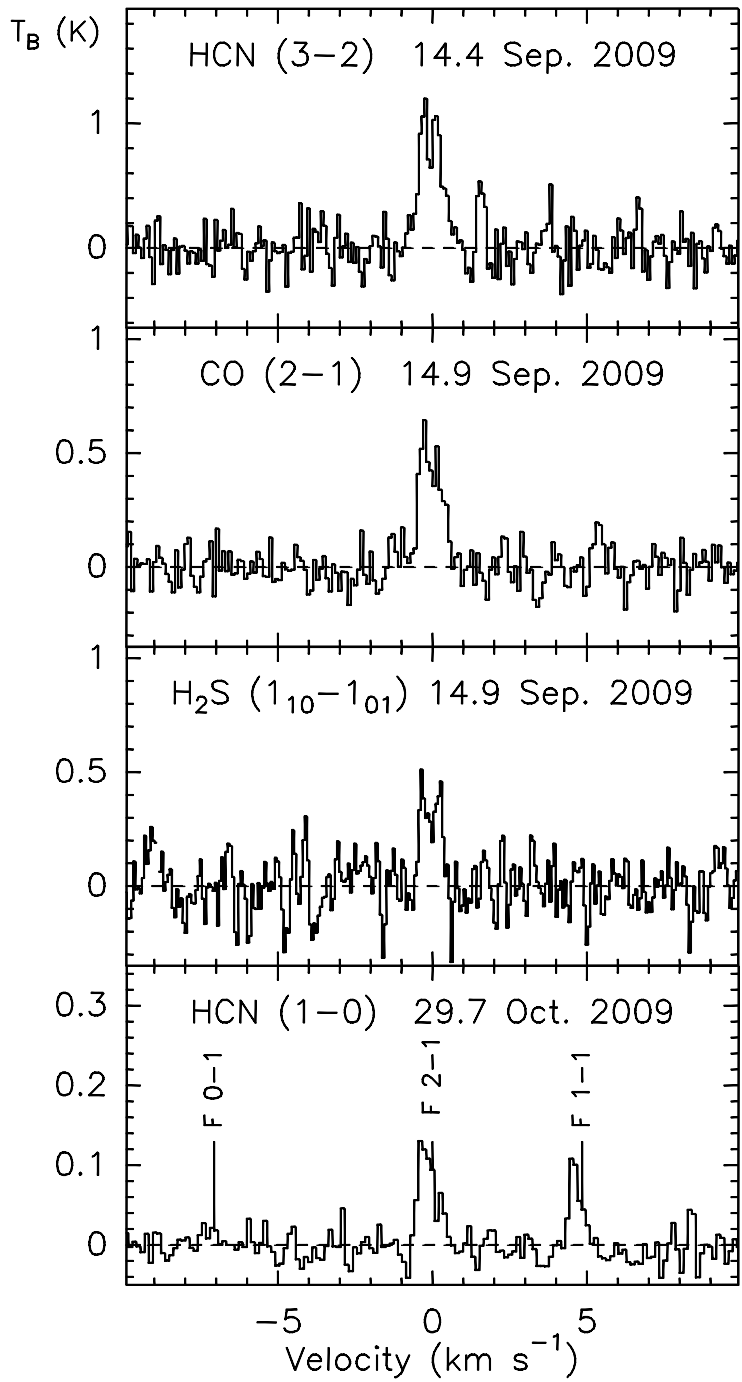}
\caption{Spectra of C/2006 W3 (Christensen) observed with the IRAM 30-m telescope.} \label{Fig:iram}
\end{figure*}
}

\onlfig{6}{
\begin{figure*}
\includegraphics[width=13cm]{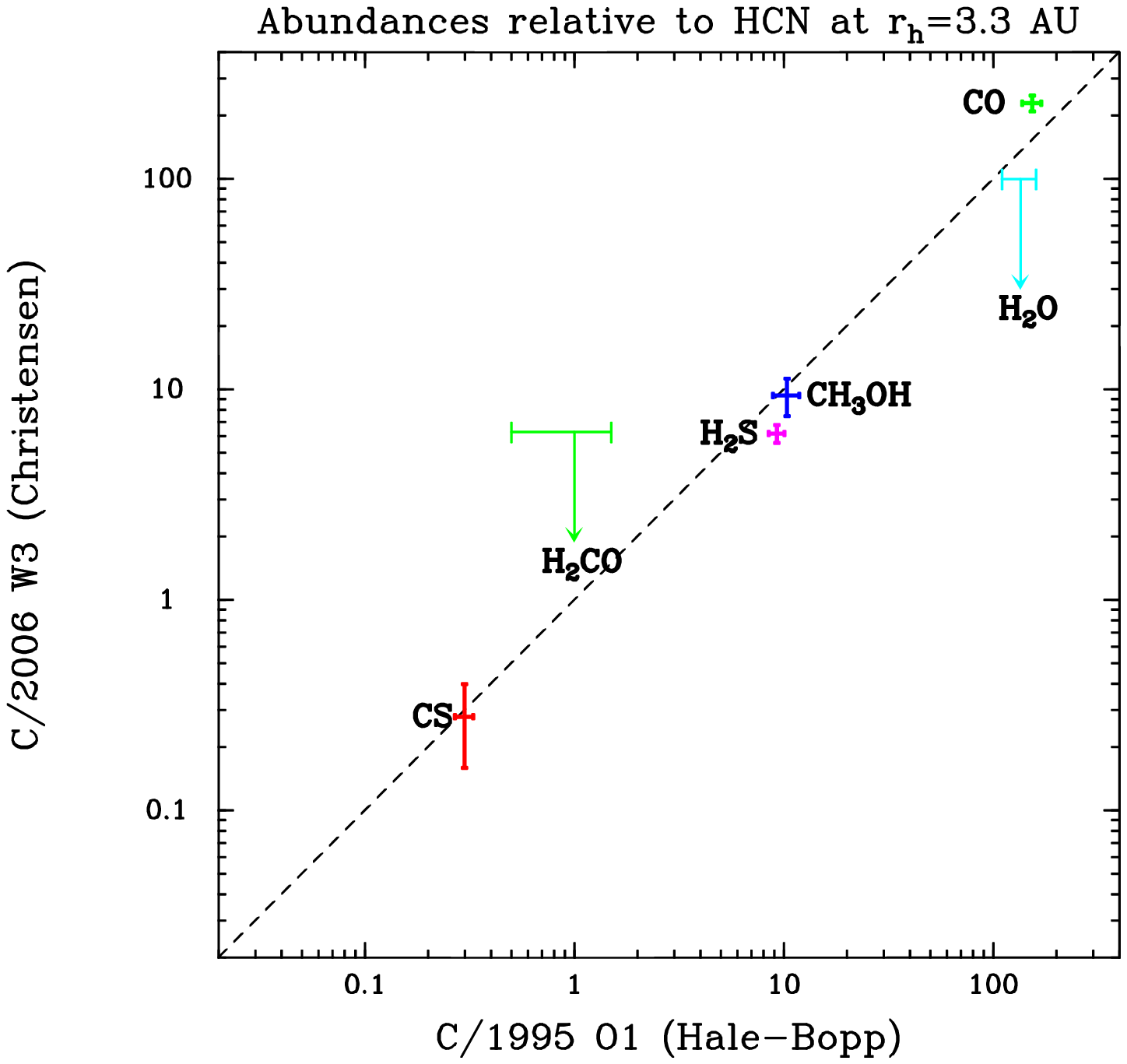}
\caption{Comparison of measured abundances relative to HCN in
comets C/2006 W3 (Christensen) and C/1995 O1 (Hale-Bopp) at 3.3 AU
from the Sun.} \label{Fig:abun}
\end{figure*}
}

The OH lines at 18 cm were observed in comet C/2006 W3
(Christensen) with the Nan\c{c}ay radio telescope
(Fig.~\ref{Fig:nancay-iram}, Table~\ref{table-I}). The methods of
observation and analysis are described in \citet{cro02}. The
observations were done pre-perihelion from February 10 to April 19 
2009 when the comet was at an average heliocentric distance
similar to that of the {\it Herschel} post-perihelion observations. An
average OH production rate $Q_{\rm OH}$ = $3.8\pm0.9\times10^{28}$ s$^{-1}$ 
is derived,  which corresponds to a water production 
rate $Q_{\rm H_2O}$ = 1.1 $\times$ $Q_{\rm OH}$ = $4.2\pm1.0\times10^{28}$ s$^{-1}$, 
assuming that water is the main source of OH radicals.

Observations with the IRAM 30-m were made on September 12--14 2009
with the recently installed EMIR receiver, completed by a short
observation on October 29 just before the {\it Herschel} observations
(Fig.~\ref{Fig:nancay-iram} and $online$ Fig.~\ref{Fig:iram},
Table~\ref{table-I}). CO, HCN (two rotational transitions), CH$_{3}$OH (six lines
around 157~GHz) and H$_{2}$S were detected. Beam sizes are between
10 and 27\arcsec, similar to PACS and SPIRE fields of view.

\section{Analysis and discussion}

The observed water rotational lines are optically thick \citep[e.g.,][]
{biv07,hart10}. To derive upper limits for the water production
rate, we use an excitation model which considers radiation
trapping, excitation by collisions with neutrals (here CO) and
electrons, and solar IR pumping \citep{biv97,biv99}. A H$_2$O-CO total
cross-section for de-excitation of 2 $\times$ 10$^{-14}$ cm$^2$ is
assumed  \citep{biv99}. The same excitation processes are considered in the
models used to analyze the molecular lines observed at the IRAM
30-m. We used a gas kinetic temperature of 18~K, derived from 
the relative intensities of the CH$_{3}$OH lines, and  
a gas expansion velocity of 0.47 km~s$^{-1}$, consistent with 
the widths of lines detected at IRAM. The largest model uncertainty 
is in the electron density profile. We used the profile
derived from the 1P/Halley in situ measurements and scaled it to the
heliocentric distance and activity of comet Christensen, as detailed in, 
e.g., \citet{ben04}. The electron density was then multiplied by a factor 
$x_{ne}$ = 0.2, constrained from observations of the 557 GHz
water line in other comets \citep{biv07,hart10}. Upper limits on the 
water production rate obtained from the 2$_{12}$--1$_{01}$ line observed with
PACS, and from the 1$_{11}$--0$_{00}$ line observed with SPIRE, are given in 
 Table~\ref{table-I}. Other observed H$_2$O lines, 
either with PACS or with SPIRE, do not improve these limits.
The higher water production rate measured pre-perihelion at
Nan\c{}cay may suggest a seasonal effect. As the field of view of
the Nan\c{}cay telescope is large (3.5\arcmin$\times$19\arcmin), another
interpretation is the detection of water sublimating from icy
grains. The $Q(\rm H_2O)$ measured at Nan\c{}cay corresponds to a
sublimation cross-section of 400 km$^2$, or a sublimating sphere
of pure ice of 8 km radius.

With a CO production rate of 3$\times$10$^{28}$ s$^{-1}$, comet Christensen 
is only four times less productive than comet Hale-Bopp at $r_h$ = 3.3 AU 
\citep{biv02}. 
The post-perihelion measurements show that the CO to H$_2$O production rate 
ratio in comet Christensen exceeds 220\%, indicating a CO-driven activity
(Table~\ref{table-I}). For comparison, CO/H$_2$O was $\sim$120\%
in comet Hale-Bopp at 3.3 AU from the Sun. When normalized to HCN,
abundance ratios HCN:CO:CH$_3$OH:H$_2$S:CS are 1:240:9:6:0.3 and
1:150:10:9:0.3 for comets Christensen and Hale-Bopp, respectively
\citep[$online$ Fig.~\ref{Fig:abun},][]{biv02}. Therefore, besides being
depleted in H$_2$O, comet Christensen is enriched in CO relative to HCN,
while other molecules have similar abundances.

The dust coma is highly condensed but clearly resolved in both
blue (B) and red (R) PACS images (Fig.~\ref{Fig:pacsmap}). The
width at half maximum of the radial profiles 
 is 9.0\arcsec~and 18.1\arcsec~ in B and R,
respectively, a factor 1.64 larger than the PSF
($\sim$5.5\arcsec~in B, 11\arcsec~in R). The B image is extended
Westward towards the Sun direction at PA = 257.5$^{\circ}$ (phase
angle = 16$^{\circ}$). This asymmetry is also seen in the R image.
HCN and CO spectral lines are similarly blue-shifted
(Table~\ref{table-I}), which indicates preferentially dayside
emission of these molecules from the nucleus, and is consistent
with the dust coma morphology. Since CO is likely the main
escaping gas \citep[CO$_2$ has been observed to be less abundant than CO
in comets][]{bock04}, dust-loading by CO gas is suggested. The enhanced CO
production towards the Sun implies sub-surface production at
depths not exceeding the thermal skin depth ($\leq$ 1 cm). The distant CO
production in comet Christensen may result from the
crystallization of amorphous water ice immediately below the
surface \citep{pria04}.

\begin{figure}[!]
\resizebox{\hsize}{!}{\includegraphics*{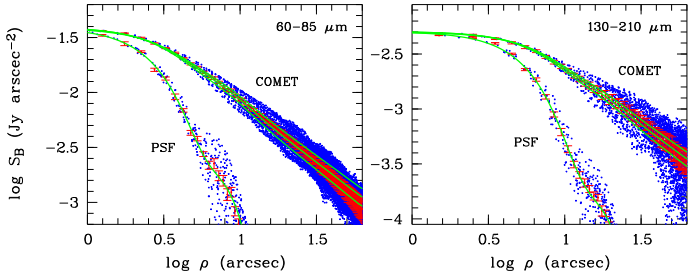}}
\caption{Comet surface brightness and re-scaled PSF (Vesta) in B 
(left box) and R (right box) bands. 
Blue dots: surface brightness for each pixel. Red
dots: mean values in 0.5\arcsec~(B) or 1.0\arcsec~(R) wide annuli,
with error bars. For the comet, the continuous green lines are radial profiles
obtained by convolving a $\rho^{-x}$ law with the PSF with, from
top to bottom, x = 1.0, 1.05, 1.1 (B) and x = 0.95, 1.0, 1.05 (R).
The green line superimposed on Vesta data is the PSF model (see text).}
\label{fig:proradB}
\end{figure}

Radial profiles in the B and R bands are presented in
Fig.~\ref{fig:proradB} for both the comet and the PSF (from Vesta
data). Although highly structured, the background has been
estimated as best as possible and subtracted.  The PSF has a
complex shape which has been fitted radially with two gaussian
profiles centered at $\rho$ = 0 and $\rho$ = 6\arcsec\ (resp.
12\arcsec) for the B (resp. R) bands.  A symmetric 2D PSF has then
been constructed and convolved with theoretical cometary surface
brightness profiles $S_{B} \propto \rho^{-x}$. One finds that the
full range of radial profiles can be reproduced with $0.8<x<1.2$.
Most of the dispersion is likely due to the PSF structure and more
specifically to its extended tri-lobe pattern \citep{lut10}. The
average surface brightness profiles can be fitted with $x = 
1.05\pm0.05$ in the B band and $x = 1.00\pm0.05$ in the R
band.  The radial dependence of the surface brightness is then
compatible with the steady-state $\rho^{-1}$ law in the full
60--210 $\mu$m spectral range.

There is then no evidence for a significant contribution of
nucleus thermal emission in the central pixels. Assuming that the
CO production rate scales proportionally to the nucleus surface
area, the size of Christensen's nucleus is estimated to be a
factor of two smaller than Hale-Bopp's nucleus size, i.e.
$D$$\sim$ 20 km for comet Christensen \citep[see][for Hale-Bopp's nucleus 
size]{alten99}.  From
the Standard Thermal Model \citep{lebo89} with the nucleus albedo set to 4\% 
 and the emissivity and beaming factor taken as unity, a
20 km diameter body would contribute to 4.5\% of the flux  measured
in the central 1\arcsec~pixel of the B image ($F_{\rm B}$ = 38.4 mJy/pixel), 
and still less (2\%) in the R image where the measured peak intensity
is  $F_{\rm R}$ = 19.5 mJy/pixel (2\arcsec-pixel). From the
B image, we estimated that $D < 26 $ km.

The Spectral Energy Distribution (SED) of the dust thermal
emission can constrain important 
properties of cometary dust, in
particular the dust size distribution and production rate
\citep{jew90}. The flux density in the SPIRE spectrum
(Fig.~\ref{Fig:spire-spectrum}) varies as $\nu^{-1.78}$. The spectral index, 
close to $\alpha$ = 2, 
indicates the presence of large grains, consistent with the
maximum ejectable dust size loaded by the CO gas
\citep[$a_{max}$ $\sim$  0.9 mm for a $D$ = 20 km  body of density
$\rho_N$ = 500 kg m$^{-3}$, with the dust density $\rho_d$ =
$\rho_N$; $a_{max}\propto D^{-3}$;][]{cri05}. We modelled the 
thermal emission of the dust coma following \citet{jew90}.
Absorption cross-sections calculated with the Mie theory were used
to compute both the temperature of the grains, solving the
equation of radiative equilibrium, and their thermal emission.
Complex refractive indices of amorphous carbon and olivine
(Mg:Fe = 50:50) \citep{edo83,dors95} were taken as broadly
representative of cometary dust. We considered a differential dust
production $Q_d$(a) as a function of size, with sizes between 0.1
$\mu$m and 0.9 mm. The
size-dependent grain velocities $v_d$(a) were computed following
\citet{cri97} assuming $D$ = 20 km, and vary  from 6 to 224 m
s$^{-1}$ ($\propto$ $a^{-0.5}$ at large sizes). We assumed a local
dust density $\propto r^{-2}$, where $r$ is the distance to 
the nucleus, consistent with the maps.   The best fit to the flux ratio
$F_{\rm B}/F_{\rm R}$ in R and B bands is obtained for 
$Q_d$(a) $\propto$ $a^{-(3.6^{+0.25}_{-0.8})}$, which yields a dust opacity 
(the ratio between the effective emitting dust cross-section and dust  
mass), of 6.8$^{+1.2}_{-2.8}$ and 5.7$^{+0.3}_{-1.3}$  m$^{2}$ kg$^{-1}$ at 450 GHz, 
and dust production rates of 850$^{+1100}_{-200}$ and 920$^{+730}_{-110}$ kg s$^{-1}$, for carbon and 
olivine grains, respectively. The whole SED between 450 and 4300 GHz is
consistently explained (Fig.~\ref{Fig:spire-spectrum}). Assuming that CO is the main gas
escaping from the nucleus, the inferred dust to gas mass
production ratio is then 0.5 to 1.4.

\section{Conclusions}
Comet Christensen was a distant comet. Nevertheless, the
continuum was clearly detected by PACS and SPIRE, providing useful 
constraints on the properties of the cometary dust. Although
water emission was not detected in this object, the limits
obtained are significant. The prospects for future cometary
studies with {\it Herschel} are thus very good.

\begin{acknowledgements}
PACS has been developed by a consortium of institutes led by MPE
(Germany) and including UVIE (Austria); KU Leuven, CSL, IMEC (Belgium);
CEA, LAM (France); MPIA (Germany); INAF-IFSI/OAA/OAP/OAT, 
LENS, SISSA (Italy); IAC (Spain). This development has been
supported by the funding agencies BMVIT (Austria), ESA-PRODEX 
(Belgium), CEA/CNES (France), DLR (Germany), ASI/INAF (Italy), and
CICYT/MCYT (Spain).
SPIRE has been developed by a consortium of institutes led by
Cardiff University (UK) and including Univ. Lethbridge (Canada);
NAOC (China); CEA, LAM (France); IFSI, Univ. Padua (Italy); IAC
(Spain); Stockholm Observatory (Sweden); Imperial College London,
RAL, UCL-MSSL, UKATC, Univ. Sussex (UK); and Caltech, JPL, NHSC, 
Univ. Colorado (USA). This development has been supported by
national funding agencies: CSA (Canada); NAOC (China); CEA, CNES,
CNRS (France); ASI (Italy); MCINN (Spain); Stockholm Observatory
(Sweden); STFC (UK); and NASA (USA).
Additional funding support for some instrument activities has been
provided by ESA.
HCSS/HSpot/HIPE are joint
developments by the {\it Herschel} Science Ground Segment Consortium,
consisting of ESA, the NASA {\it Herschel} Science Center, and the HIFI,
PACS and SPIRE consortia.
IRAM is an international institute
co-funded by CNRS, France, MPG, 
Germany, and IGN, Spain.
The Nan\c{c}ay radio observatory is cofunded by CNRS, Observatoire
de Paris, and the R\'egion Centre (France).
%
%
%
%
%
%
D.B.-M. thanks M.A.T. Groenewegen and D. Ladjal for support in PACS data analysis,
and V. Zakharov for useful discussions on gas and dust dynamics.

\end{acknowledgements}


\end{document}